\documentclass[a4paper,fleqn,usenatbib]{mnras}

\usepackage{txfonts}

\usepackage[T1]{fontenc}
\usepackage{ae,aecompl}

\usepackage{psfig}   
\usepackage{graphicx}
\usepackage{amssymb}
\usepackage{subfigure}

\title[Enlarging habitable zones]{Enlarging habitable zones around binary stars in hostile environments}

\author[B. A. Wootton and R. J. Parker]{Bethany A. Wootton and Richard  J. Parker\thanks{E-mail: R.Parker@sheffield.ac.uk}\thanks{Royal Society Dorothy Hodgkin Fellow} \vspace*{0.1cm}\\
Department of Physics and Astronomy, The University of Sheffield, Hicks Building, Hounsfield Road, Sheffield, S3 7RH, UK}

\begin{document}

\date{}
                             
\pagerange{\pageref{firstpage}--\pageref{lastpage}} \pubyear{2018}

\maketitle

\label{firstpage}

\begin{abstract}
Habitable zones are regions around stars where large bodies of liquid water can be sustained on a planet or satellite. As many stars form in binary systems with non-zero eccentricity, the habitable zones around the component stars of the binary can overlap and be enlarged when the two stars are at periastron (and less often when the stars are at apastron). We perform $N$-body simulations of the evolution of dense star-forming regions and show that binary systems where the component stars originally have distinct habitable zones can undergo interactions that push the stars closer together, causing the habitable zones to merge and become enlarged. Occasionally, overlapping habitable zones can occur if the component stars move further apart, but the binary becomes more eccentric. Enlargement of habitable zones  happens to 1--2 binaries from an average initial total of 352 in each simulated star-forming region, and demonstrates that dense star-forming regions are not always hostile environments for planet formation and evolution.
\end{abstract}

\begin{keywords}   
open clusters and associations: general -- planets and satellites: terrestrial planets --  astrobiology
\end{keywords}

\section{Introduction}

One of the main drivers of exoplanet research is to find an Earth-like planet that orbits within the habitable zone of the planet's host star. The exact definition and boundaries of habitable zones vary, but it is generally accepted that the habitable zone is the location where water can exist in liquid form \citep{Kasting93,Ramirez18}, as on Earth.

The calculation of the habitable zone for a single star like the Sun is relatively straightforward, with the flux (i.e. effective temperature) of the host star the main variable. To a lesser extent, the composition of the atmosphere, and the amount of cloud cover on a planet are also determining factors \citep{Kasting93}. Habitable zones around single stars therefore only change if the luminosity of the host star changes (e.g. due to stellar evolution from the pre- to Main Sequence, and then once the star moves off the Main Sequence toward the end of its life), or the atmosphere of any planet changes drastically.

However, most stars are not born as singles, but rather in binary or higher-order multiple systems where the semimajor axis can be anywhere between $\sim 0.01 - 10^4$\,au \citep{Duchene13b}. The binary fraction of solar-mass primary stars in the Sun's local Galactic environment is just less than 50\,per cent \citep{Raghavan10}, though lower-mass primaries (e.g. $m_p < 0.5\,$M$_\odot$) likely have a lower binary fraction \citep[30 -- 40\,per cent,][]{Bergfors10,Janson12,Ward-Duong15} whereas higher mass primary stars (>1.5\,M$_\odot$) have a higher binary fraction \citep[increasing towards 100\,per cent for the most massive stars,][]{DeRosa14,Sana13}. 

The binary fraction is more difficult to measure in Giant Molecular Clouds where stars (and planets)  are forming, and the values in the Galactic disc are thought to be a lower-limit to the initial fraction of stars that are found in binaries. In star-forming regions the stellar density can exceed the value in the Galactic  disc by several orders of magnitude \citep{King12a}, and at these densities ($>$100\,M$_\odot$\,pc$^{-3}$) dynamical interactions can change the initial orbits of binary systems, or destroy the system altogether to produce two single stars. Conversely, it is difficult to form all but the most extreme wide binary ($a_{\rm bin}>1000$\,au) systems through dynamical interactions \citep{Kouwenhoven10,Moeckel10}, and systems with semimajor axes lower than these values are usually primordial.

Observational and theoretical work has shown that binary stars do not preclude planet formation. Radial velocity surveys \citep[e.g.][]{Raghavan06,Bonavita07} have found planets orbiting the primary component of binary systems \citep[so-called `S-type', or satellite orbits,][]{Dvorak86} and more recently the \emph{Kepler} mission has shown that planets can orbit both components of a binary system \citep{Doyle11} in a so-called `P-type', or primary orbit \citep{Dvorak86}. 

Furthermore, dynamical stability calculations have shown that planets can exist on long-lived orbits in S-type systems, usually if they orbit their host star at a distance of $<0.3a_{\rm bin}$ and the eccentricity of the binary is not too high  \citep[$e_{\rm bin} < 0.9$,][]{Holman99}. In this situation, each component of the binary could feasibly host planets, and each component of the binary will have its own habitable zone, unless the component stars are so close that their habitable zones merge. This varies depending on the masses of (and hence fluxes from) the component stars of the binary, but for a mildly eccentric ($e = 0.5$) binary with semimajor axis $\sim 5$\,au it is possible that the binary habitable zone will merge and be larger than if the two stars were single or widely separated \citep{Kaltenegger13,Cuntz14,Jaime14}.

This then raises the interesting possibility that the orbit of a binary star system can be altered due to dynamical interactions in its birth star-forming region to the extent that the previously separate habitable zones merge, and the enhanced incident flux from both stars enlarges the width of the habitable zone around both stars. In star-forming regions, binaries with a binding energy higher than the local average energy of passing stars will not break apart, and are also likely to have their semimajor axis decreased or ``hardened". (A binary with binding energy less than the surrounding stars is ``soft" and will likely break apart or increase its semimajor axis after an interaction -- \citealp{Heggie75,Hills75a}.) The boundary between the hard and soft regimes depends on the stellar density, interaction rate and local velocity dispersion in a star-forming region, but the latter two variables are more or less constant. 

In very dense star-forming regions  ($>$1000M$_\odot$\,pc$^{-3}$) the hard-soft boundary is several au, which is the regime in which the habitable zones of main sequence stars in binaries can overlap. In this Letter, we determine how often binary stars are hardened in dense star-forming regions such as the Orion Nebula Cluster (ONC), and whether this causes the habitable zones around the component stars to merge and become enlarged. We describe our simulations of the evolution of star-forming regions and outline our method to calculate the habitable zones in Section 2. We present our results in Section 3 and we conclude in Section 4.

\section{Method}

\subsection{$N$-body simulations}

We set up $N$-body simulations of star-forming regions with spatial and kinematic substructure in an attempt to mimic observations that indicate that young star-forming regions appear spatially clumpy and filamentary. The local velocity dispersions are small, meaning that stellar velocities are correlated on local scales. We create substructured star-forming regions containing 1500 stars arranged in a fractal distribution according to the prescription in \citet{Goodwin04a}, with fractal dimension $D = 1.6$, resulting in a high degree of spatial and kinematic substructure.

The star-forming regions have initial radii of 1\,pc, and \emph{primary} masses drawn from a \citet{Maschberger13} Initial Mass Function (IMF), which has a probability distribution of the form
\begin{equation}
p(m) \propto \left(\frac{m}{\mu}\right)^{-\alpha}\left(1 + \left(\frac{m}{\mu}\right)^{1 - \alpha}\right)^{-\beta},
\end{equation}
where $\mu = 0.2$\,M$_\odot$ is the average stellar mass, $\alpha = 2.3$ is the \citet{Salpeter55} power-law exponent for higher mass stars, and $\beta = 1.4$ describes the slope of the slope of the IMF for low-mass objects. We sample this distribution in the mass range 0.01 -- 50\,M$_\odot$. The average local stellar density in these simulations is $\simeq 10^4$M$_\odot$\,pc$^{-3}$, consistent with inferred initial densities of some star-forming regions such as the ONC \citep{Marks12,Parker14e}.

We determine whether a star is the primary component of a binary system depending on its mass. Observations of binary stars where the primary component  is roughly Solar mass in the local Galactic  neighbourhood suggest a binary fraction of $f_{\rm bin} = 0.46$, where 
\begin{equation}
f_{\rm bin} = \frac{B}{S + B}
\end{equation}
and $S$ and $B$ are the number of single and binary systems, respectively. This fraction decreases for lower-mass primary systems and increases for higher-mass systems. The distribution of orbital semimajor axes of Solar-type primaries can be fit with a log-normal distribution with mean 50\,au and a variance log$_{\rm 10} \sigma_a = 1.68$. The distribution of mass ratios $q = m_s/m_p$ is observed to be flat in the field \citep{Reggiani11a}  and the distribution of eccentricities is also flat for binaries with a semimajor axis greater than 0.1\,au \citep{Raghavan10}, with systems on shorter orbits tending to zero eccentricity. We draw mass ratios and eccentricities from these distributions and then draw semimajor axes according to the primary mass  (we summarise the adopted binary fraction, peak of the semimajor axis distribution and variance of this distribution for a given primary mass range in Table~\ref{bin_props}). 

\begin{table}
  \caption[bf]{Summary of the adopted binary fraction and semimajor axis distribution as a function of primary mass for stars in the Galactic field which we use to set up the binary populations in our star-forming regions.}
  \begin{center}
    \begin{tabular}{|c|c|c|c|c|}
      \hline
Mass range & $f_{\rm bin}$ & $\tilde{a}$ & log$_{\rm 10} \sigma_a$ & refs. \\	
      \hline
      $0.01 < m_p \leq 0.08$ & 0.15 & 4\,au & 0.4 & (1) \\
$0.08 < m_p \leq 0.5$ & 0.34 & 16\,au & 0.8 & (2),(3),(4) \\
$0.5 < m_p \leq 1.5$ & 0.46 & 50\,au & 1.68 & (5) \\  
$1.5 < m_p \leq 3$ & 0.48 & 230\,au & 0.79 & (6) \\  
      \hline
    \end{tabular}
  \end{center}
  \label{bin_props}
  References: (1) \citet{Burgasser07}; (2) \citet{Bergfors10}; (3) \citet{Janson12}; (4) \citet{Ward-Duong15}; (5) \citet{Raghavan10}; (6) \citet{DeRosa14}
\end{table}

We then randomly distribute our primordial binary systems and remaining single stars within the fractal distributions. We use the \texttt{kira} integrator \citep{Zwart01}, a 4$^{th}$-order Hermite scheme $N$-body integrator within the \texttt{Starlab}  environment to evolve the simulations for 10\,Myr which is long enough for most dynamical processing to act on the binary systems in dense stellar environments. We do not include internal stellar or binary evolution as we do not expect theses processes to be important for stars with $m < 3$M$_\odot$ in the 10\,Myr timeframe of the simulations.

\subsection{Calculating the habitable zones in binaries}

In our analysis we take the primary and secondary masses of the binary system, as well as the semimajor axis and eccentricity, to calculate the extent of the habitable zones before dynamical evolution ($t$ = 0\,Myr) and at the end-point of the simulation ($t$ = 10\,Myr). We use the mass--luminosity--temperature relations for Main Sequence stars in \citet{Cox00}, where $L/L_\odot = \left(M/M_\odot\right)^{3.5}$ to derive an effective temperature for each star in a given binary system. We do not use pre-main sequence temperatures and luminosities because we are interested in the habitability of the star once it reaches the main sequence, and how different this habitability would have been had the binary system in question not had its orbit altered by a dynamical interaction in the birth star-forming region.

We follow the method adopted by \citet{Kaltenegger13} to calculate the narrow and empirical habitable zones around each component star in the binary. A more generalised version of these calculations was derived by \citet{Cuntz14,Cuntz15} -- see also \citet{Jaime14}. For individual stars, the position of the outer or inner boundary of the habitable zone is given by
\begin{equation}
l_{\rm x-Star} = l_{\rm x-\odot}\left[\frac{L_{\rm Star}/L_\odot}{1 + \alpha_{\rm x}(T_i)l_{\rm x-\odot}^2}\right]^{1/2},
\end{equation}
where $l_{\rm x-\odot}$ is the inner or outer boundary of the Sun's habitable zone, $L_{\rm Star}$ is the luminosity of the star, $L_\odot$ the luminosity of the Sun. $\alpha_{\rm x}(T_i)$ is given by
\begin{equation}
\alpha_{\rm x}(T_i) = a_{\rm x}T_i + b_{\rm x}T_i^2 + c_{\rm x}T_i^3 + d_{\rm x}T_i^4 ,
\end{equation}
where
\begin{equation}
T_i({\rm K}) = T_{\rm Star}({\rm K}) - 5870,
\end{equation}
and  $a_{\rm x}$, $b_{\rm x}$, $c_{\rm x}$  and $d_{\rm x}$ are coefficients which depend on the type of habitable zone under consideration and are found in \citet{Kopparapu13} and also summarised in table~1 in \citet{Kaltenegger13}. The subscript $_i$ refers to either primary component $_p$ or secondary component $_s$.


\begin{figure*}
  \begin{center}
\setlength{\subfigcapskip}{10pt}
\hspace*{-1.0cm}\subfigure[]{\label{change_bin-a}\rotatebox{270}{\includegraphics[scale=0.28]{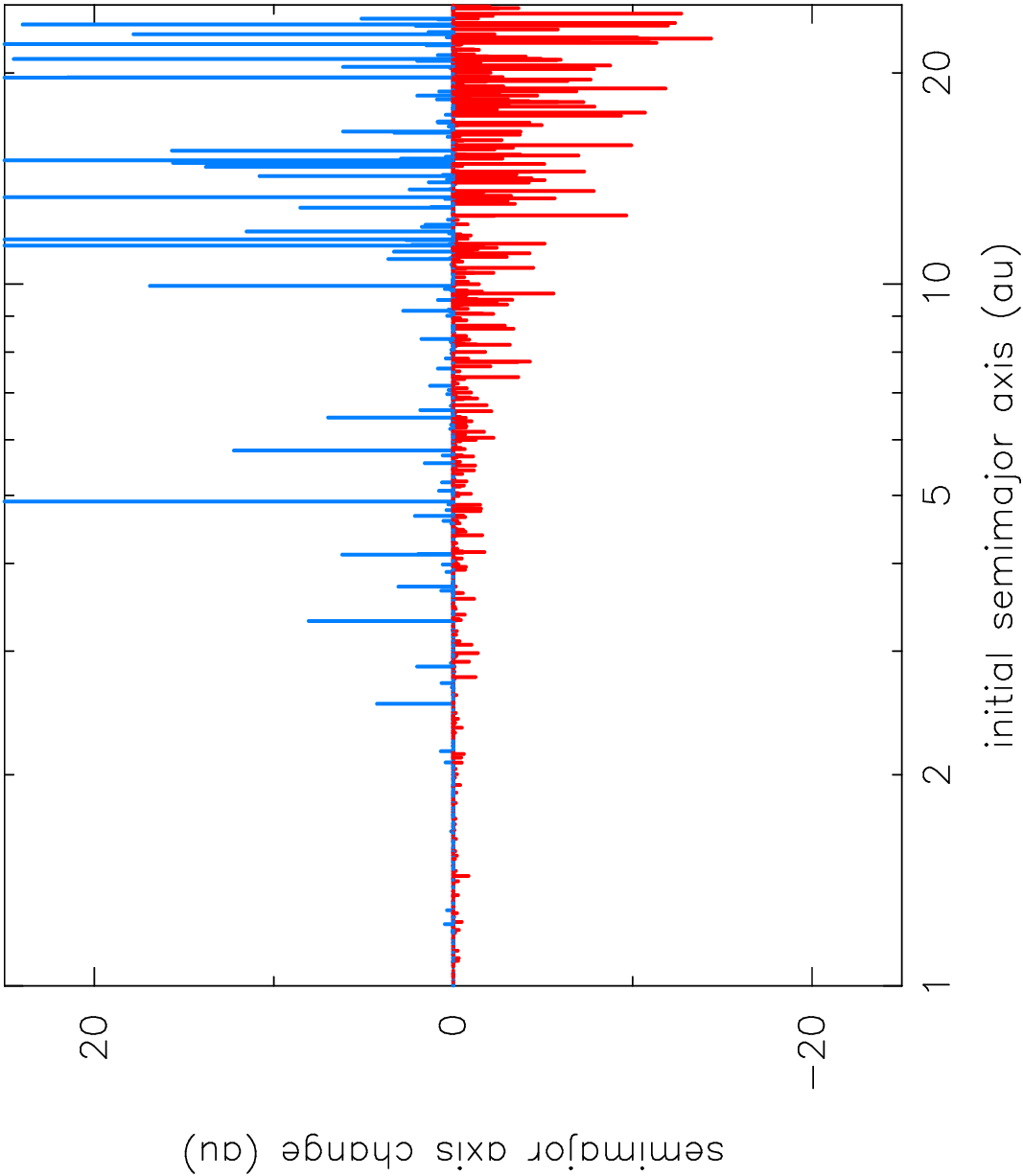}}}
\hspace*{0.3cm}
\subfigure[]{\label{change_bin-b}\rotatebox{270}{\includegraphics[scale=0.28]{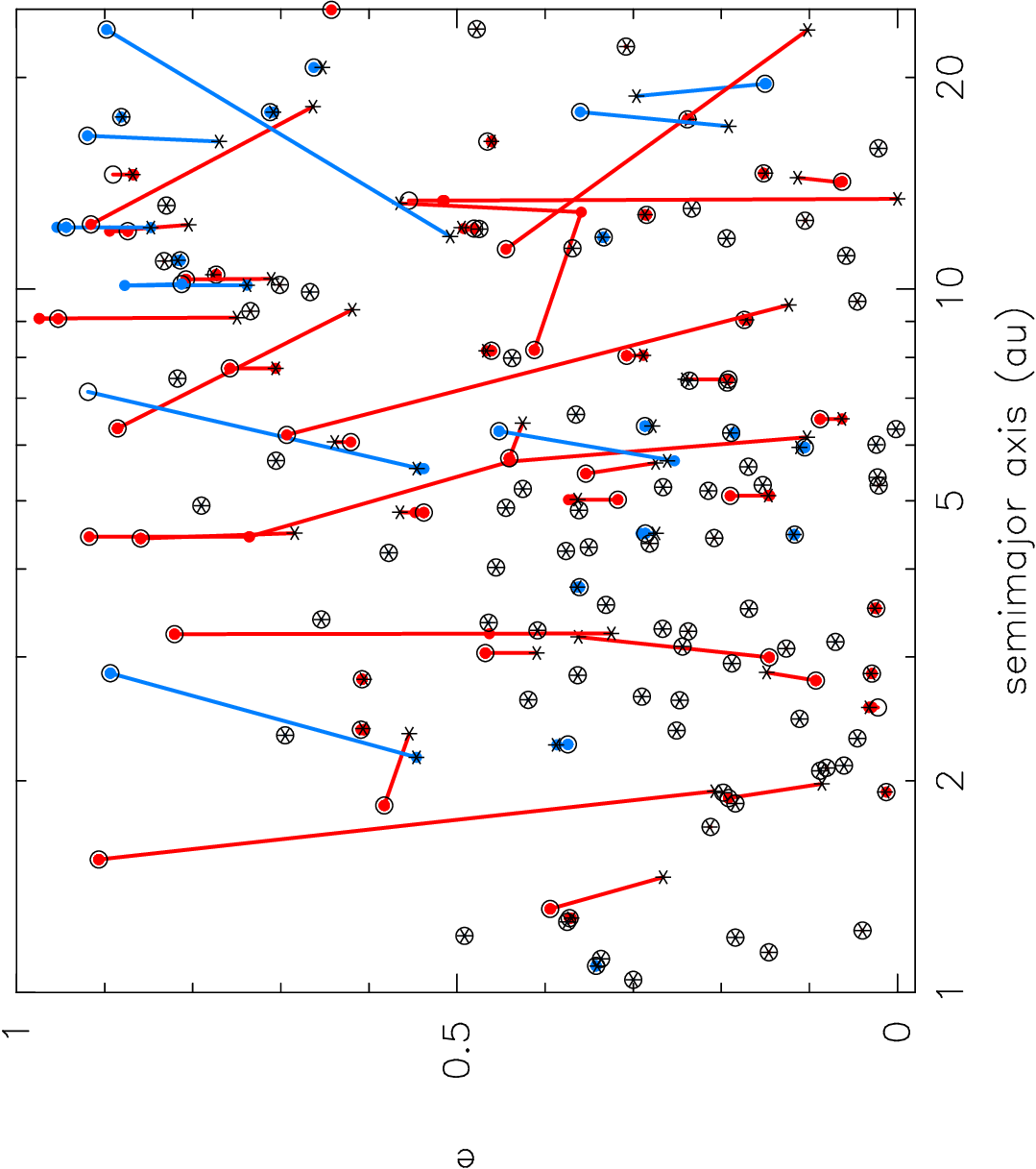}}}
\hspace*{0.3cm}\subfigure[]{\label{change_bin-c}\rotatebox{270}{\includegraphics[scale=0.28]{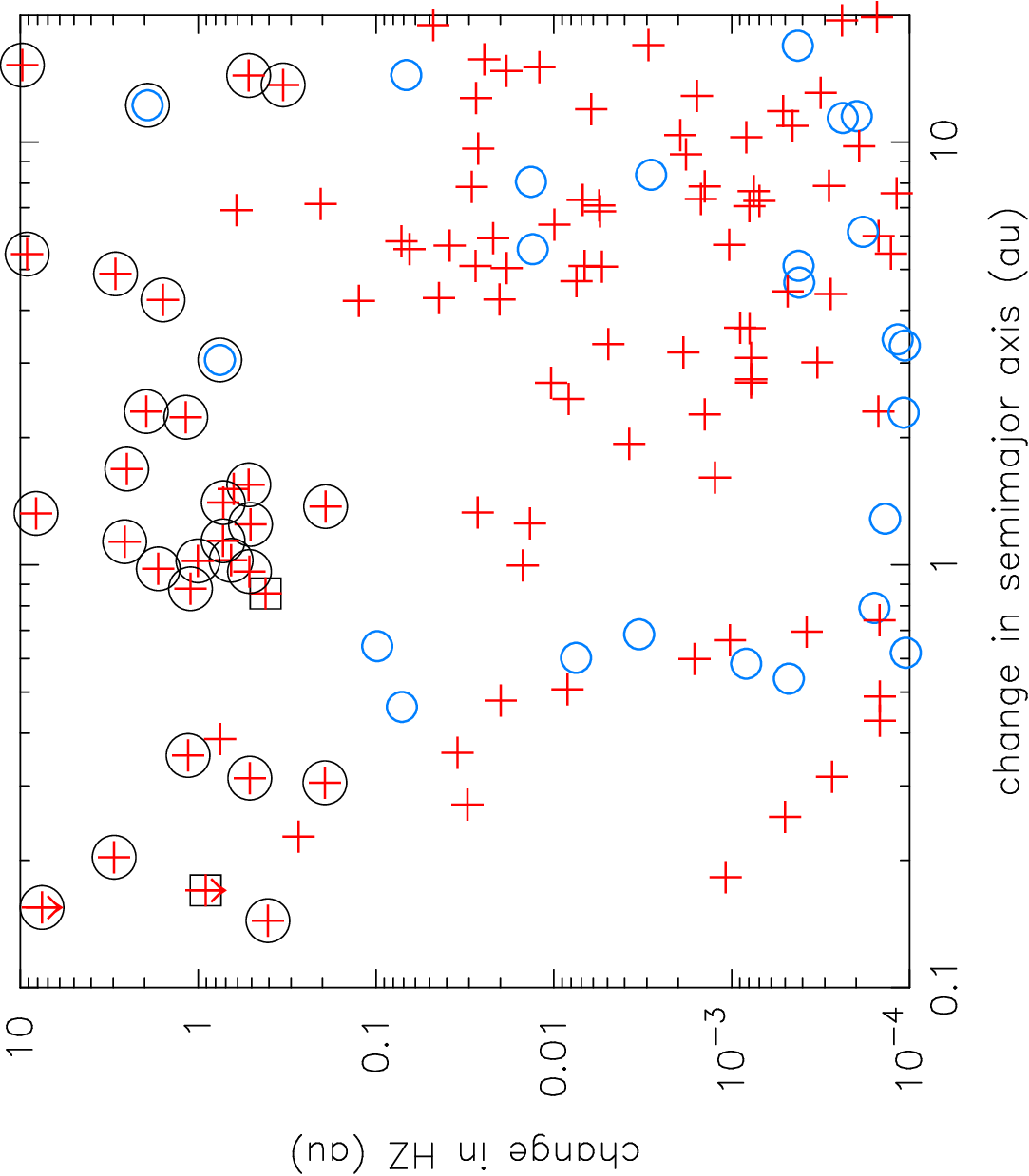}}}  
\caption[bf]{Changes to the orbital parameters of close binary star systems due to dynamical interactions in star-forming regions. In panel (a) we show the change in semimajor axis for binary stars against their initial semimajor axis. Binaries whose semimajor axes decrease (harden) due to dynamical interactions are shown by the red lines, and those whose semimajor axes increase (softens) are shown by the blue lines. In panel (b) we show eccentricity against semimajor axis. The asterisks represent the initial ($t$ = 0\,Myr) values, and the circles represent the final ($t$ = 10\,Myr) values. Filled circles are plotted at 0.1, 1, 2 and 5\,Myr.  As in panel (a), the line joining the initial and final values is red if the binary is hardened, and blue if it is softened. In panel (c) we show the change in the size of the habitable zone as a function of the change in semimajor axis. Red crosses indicate systems that are hardened, whereas blue circles indicate softened systems. Systems whose habitable zones overlap at periastron following dynamical interactions are plotted within a black circle, and systems whose habitable zones overlap at apastron are plotted within a black square. Most habitable zones are enlarged by interactions, but in two of our binaries the interaction causes the habitable zone to shrink (indicated by the downward arrow markers). }
\label{change_bin}
  \end{center}
\end{figure*}

\begin{figure*}
  \begin{center}
\setlength{\subfigcapskip}{10pt}
\hspace*{-1.5cm}\subfigure[0\,Myr]{\label{HZ_change-a}\rotatebox{270}{\includegraphics[scale=0.35]{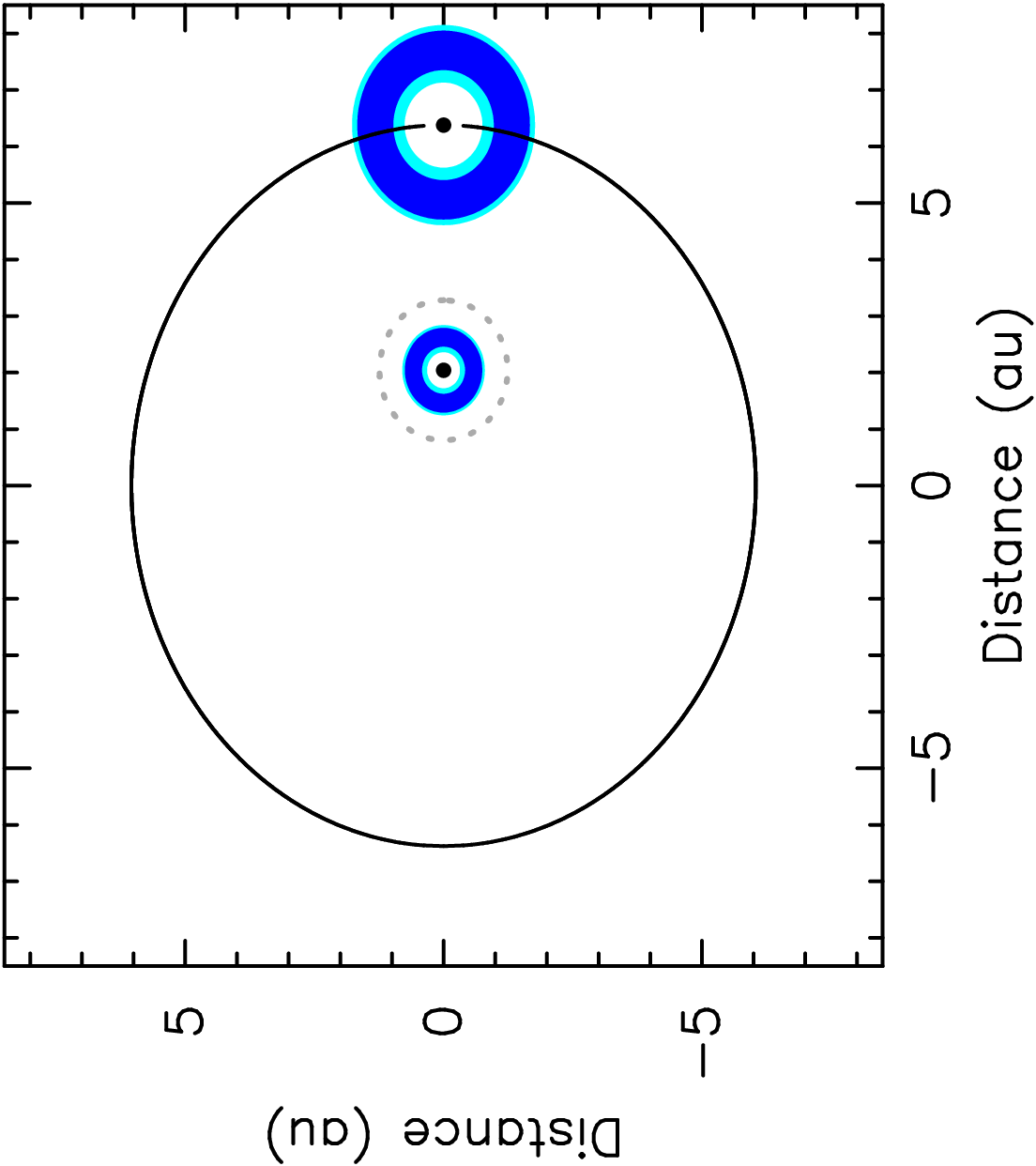}}}
\hspace*{0.3cm} 
\subfigure[10\,Myr]{\label{HZ_change-b}\rotatebox{270}{\includegraphics[scale=0.35]{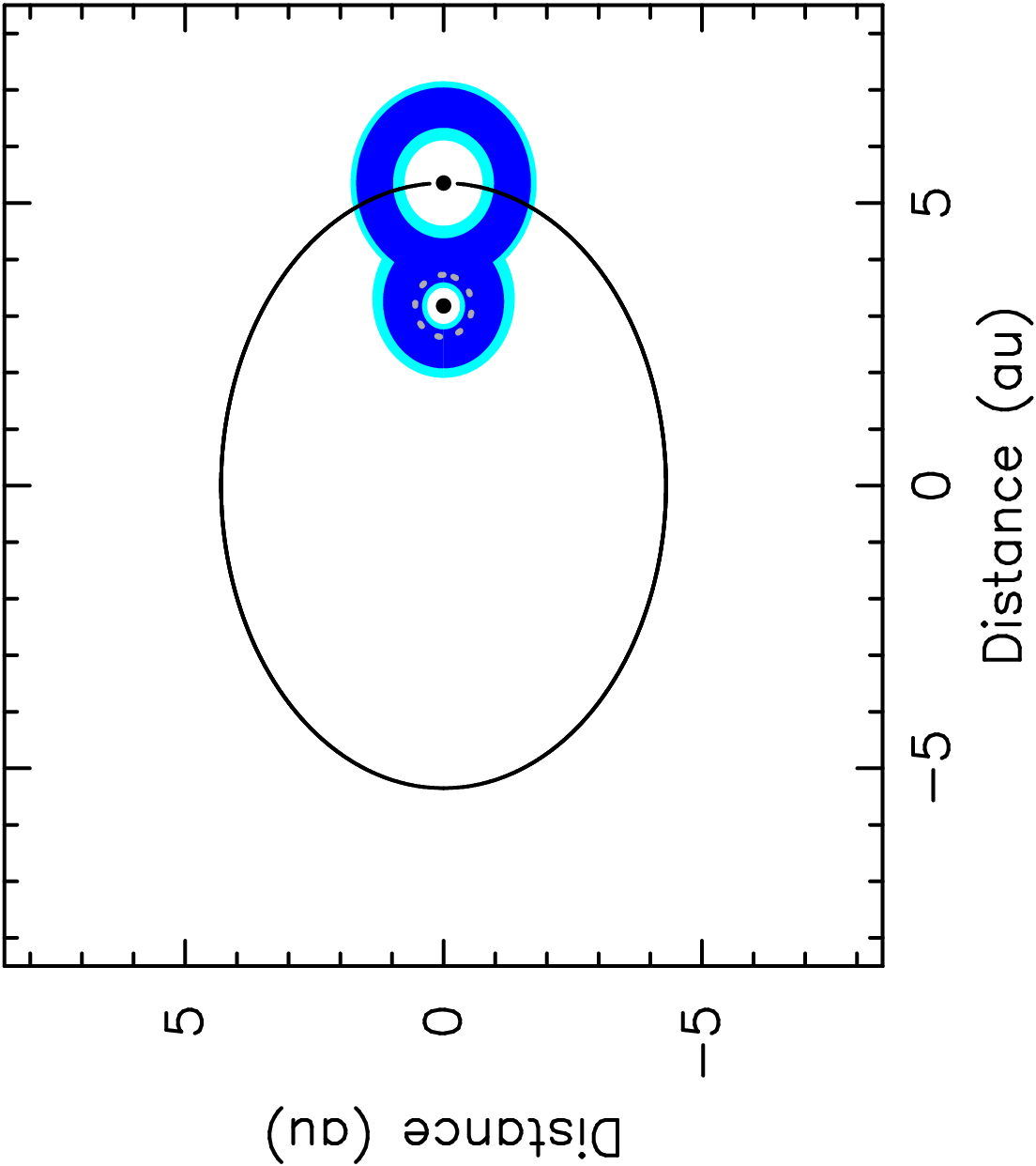}}}
\caption[bf]{The merging and enlargement of the habitable zones around the two component stars in a binary system due to the binary's hardening as a result of dynamical interactions in a star-forming region. The frame of reference is the lower-mass secondary component star, which in this example is 0.63\,M$_\odot$, and the primary component has a mass of 0.99\,M$_\odot$. The narrow habitable zone is shown by the dark blue shading, and the empirical habitable zone is shown by the cyan shading. The orbit of the binary system is shown by the solid ellipse. In panel (a) we show the binary at 0\,Myr, where the semimajor axis is 6.4\,au, eccentricity $e = 0.32$ and each star has its own distinct habitable zone. In panel (b) we show the system at 10\,Myr after it has undergone a hardening interaction to $a = 5.4$\,au and $e = 0.59$\,au. The habitable zones are now enlarged (especially that of the secondary star) and have merged. The dotted grey lines indicate the maximum semimajor axis a planet can have and remain stable within the binary system according to the criteria in \citet{Holman99} for planets on circular orbits.}
\label{HZ_change}
  \end{center}
\end{figure*}

We consider a narrow habitable zone, which varies depending on the Greenhouse Effect from CO$_2$ with boundary values for the Sun of $l_{\rm x-\odot} = 0.97$\,au for a `Runaway Greenhouse' and $l_{\rm x-\odot} = 1.67$\,au for a `Maximum Greenhouse' \citep{Kopparapu13,Kaltenegger13}. We also consider an empirical habitable zone, where $l_{\rm x-\odot} = 0.75$\,au for a `Recent Venus' and  $l_{\rm x-\odot} = 1.77$\,au for an `Early Mars'; both planets show no evidence for liquid water on their surfaces at these times, indicating they would not be within the habitable zone \citep{Kopparapu13}.

We then calculate the spectral weight factors, $W_i(f, T_i)$, which determine the relative contribution of the flux from each component of the binary star in determining the habitable zone: 
\begin{equation}
W_i(f, T_i) = \left[1 + \alpha_{\rm x}(T_i)l_{\rm x-\odot}^2 \right]^{-1},
\end{equation}
which depend on the assumed cloud cover on the planet in the habitable zone $f$ and the effective temperature of the star $T_i$. 

We combine these equations to determine the inner and outer boundaries of the narrow habitable zone, and the empirical habitable zone in the binary system, $l_{\rm x-bin} $ using the following equation
\begin{equation}
l_{\rm x-bin} = \left[\frac{W_pL_p}{L_\odot/l_{\rm x-\odot}^2 - W_sL_s/r_{\rm pl-s}^2} \right]^{1/2}.
\end{equation}
Here, $W_p$ and $W_s$ are the spectral weight factors for the primary and secondary, $L_p$ and $L_s$ are the luminosities of the component stars and $r_{\rm pl-s}$ is the position of the planet with respect to the secondary component star (recall that we are interested in how the habitable zone of the lower-mass secondary star may be enlarged, so we use the distance from the hypothetical planet to the secondary star). If the binary is eccentric ($e > 0$), $l_{\rm x-bin}$ will change as the two component stars orbit each other.

\section{Results}

We first determine the number of hard, or close, binary systems that have their orbital semimajor axis significantly hardened during dynamical interactions in our simulated star-forming regions and examine the distribution of these altered systems. In Fig.~\ref{change_bin-a} we show the change in semimajor axis of binaries against their \emph{initial} semimajor axis. Systems that are hardened are shown in red, whereas softened systems -- those whose semimajor axis increases due to a dynamical interaction -- are shown in blue. In the following we will focus on binaries with semimajor axes that are typically in the range 1 -- 20\,au as these are systems that can (a) host stable planets in their habitable zones and (b) could be altered such that the habitable zones of one or both stars could be enlarged. However, we have performed our calculations for \emph{all} surviving binaries in our star-forming regions. 

Clearly, significantly more binary systems in this range are hardened to smaller semimajor axes than are softened to larger semimajor axes.  Indeed, for every one binary that has its semimajor axis softened by more than 10\,per cent of the initial value, two binaries are hardened by more than 10\,per cent. When binary stars are subject to dynamical processing it is often the orbital eccentricity that changes the most. This is apparent in Fig.~\ref{change_bin-b}, where we show the change in eccentricity \emph{and} semimajor axis in one simulation. The initial $t$ = 0\,Myr values are shown by the asterisks and the final $t$~=~10\,Myr values are shown by the circles. We also show intermediate values at 0.1, 1, 2 and 5\,Myr by the filled symbols; however because these systems are dynamically hard, most binaries experience only one perturbative interaction, and this interaction usually occurs within the first 0.1\,Myr when the star-forming region is at its densest. The coloured lines denote whether the binary is hardened or softened. Most systems that are dynamically hardened have their eccentricities increased, which results in the two components of the binary becoming even closer at periastron.


In Fig.~\ref{change_bin-c} we show the change in the size of the habitable zone for binaries that are hardened or softened in \emph{all} simulations, as a function of the change in the semimajor axis of the binary. From a total of 7032 binaries in twenty realisations of the same star-forming region (352 per region), 4745 survive (237 per region) and of these, 354 are hardened or softened so that the habitable zone of the binary increases (18 per region).

We then consider systems where the habitable zones are enlarged such that the habitable zones around the binary components overlap. Across our ensemble of simulations, we find that 31 systems (i.e. 1--2 per region, or 0.44\,per cent of the total number of initial binaries [0.65\,per cent of the surviving binaries]) are dynamically altered such that their habitable zones expand \emph{and} overlap. Most of these systems only overlap at periastron (denoted by the symbols within the  black circles in Fig.~\ref{change_bin-c}), but we found two systems whose habitable zones even overlap at apastron (the symbols within the black squares). In the majority of systems, the habitable zone increases in size. Across all of our simulations, we find two systems where the habitable zone decreases in size; these systems are shown by the downward arrow symbols in Fig.~\ref{change_bin-c}.


We show one example of an overlapping habitable zone in Fig.~\ref{HZ_change}, and typically one to two binaries in each star forming region undergo significant  habitable zone enlargement. The binary system shown in Fig.~\ref{HZ_change} consists of a primary star of mass $m_p = 0.99$M$_\odot$ and a secondary star of mass $m_s = 0.63$M$_\odot$, initially on an orbit with semimajor axis $a = 6.4$\,au and an orbital eccentricity of $e = 0.32$.  The respective effective temperatures of these two stars are $T_{{\rm eff},p} = 5780$K and $T_{{\rm eff},s} = 4410$K. The binary then experiences an interaction that hardens the system to a final semimajor axis of $a = 5.4$\,au, but which also increases the eccentricity to  $e = 0.59$. 

Before this interaction, the binary component stars each have distinct, separate habitable zones, with the lower-mass component having a noticeable smaller habitable zone than the higher-mass (primary) component (Fig.~\ref{HZ_change-a}). Following the interaction, when the binary is at periastron, the habitable zones are enlarged and also merge together (Fig.~\ref{HZ_change-a}). 

We check the dynamical stability of the binary system before and after the interaction that causes the habitable zones to merge, using the criterion derived  by \citet{Holman99} for `S-type' orbits. For the system shown in Fig.~\ref{HZ_change}, a planet will be stable if it is within 1.24\,au of the star before the interaction, which decreases to 0.55\,au following the interaction. The narrow habitable zone for the secondary star before the interaction ranges between  0.32 -- 0.79\,au, and is then 0.46 -- 1.26\,au afterwards, which means a planet on an orbit between 0.46 and 0.55\,au would both be stable and orbit in the enlarged habitable zone. These stability criteria are shown in Fig.~\ref{HZ_change} by the dotted grey lines (planets will be stable if they orbit interior to these lines).

We determine the \citet{Holman99} stability criterion for each binary where the habitable zones are enlarged, and find that for 20\,per cent of these systems a planet would only be stable if it lay interior to the habitable zone around either star. However, we have assumed that the planets are on circular orbits, whereas planets may be on eccentric orbits if they had already formed before the binary underwent its perturbative interaction. Because most of the hardening interactions occur in the first 0.1\,Myr, a planet that is forced onto an eccentric orbit by a perturbative interaction may subsequently be circularised due to interactions with the remaning protoplanetary disc.





\section{Discussion and conclusions}

 We have shown that dynamical interactions in dense star-forming regions can dynamically alter the orbits of relatively close ($\sim$5 -- 10\,au) binary stars, pushing the two component stars closer together and often increasing the eccentricity of the system. If the binary orbit is sufficiently `hardened', i.e.\,\,to around 5\,au, the habitable zone around a low-mass secondary star ($\sim$0.5\,M$_\odot$) can be enlarged due to the increase in flux from the (typically) solar-mass primary star. 

 Depending on the orbit and the individual masses of the component star, the habitable zones overlap and are enlarged when the binary is at periastron (i.e. the stars are closest). Almost all of our systems do not have overlapping habitable zones at apastron, but the habitable zone of the lower mass star is still enlarged by the flux from the higher mass star. It is worth emphasising that any planet in the enlarged habitable zone would experience periods of lower habitability when the component stars were widely separated, unless the atmosphere of the planet can act as a buffer for the flux incident at periastron \citep{Kaltenegger13}. When the habitable zones do overlap at apastron, any planets would spend more time in the habitable zone.

 We assume that the planets are on circular orbits, but they would likely be pushed onto eccentric orbits if they formed \emph{before} the interaction that hardened the binary's orbit. We are unable to quantify the number of systems that this could occur in, but we note that in such a scenario eccentric planets could pass in and out of the habitable zone.

 Dense star-forming regions of the type we model here are usually considered to be extremely hostile to planet formation. Extreme and far-ultraviolet (EUV and FUV) radiation fields generated by OB-type stars have been shown to photoevaporate the gas content of protoplanetary discs in dense star-forming regions \citep{Adams04,Scally01}. We would therefore expect the discs from which any planets form to be primarily composed of  dust and unlikely to form gas or ice giants. However, given the closeness of the component stars in the binary, we would anyway expect that the planetary system would consist solely of terrestrial planets \citep{Haworth18a}.

We also note that we have assumed that the protoplanetary discs are circumprimary and circumsecondary, whereas in close binaries they may be circumbinary.  

Interactions that would alter the orbit of binary systems would also likely disrupt the orbits of fledgling planetary systems \citep[e.g.][]{Parker12a,Cai17a}, and planets orbiting binary stars would likely be most susceptible due to a higher cross section for collisions \citep{Adams06}. However, in our simulations the majority of the perturbative interactions occur in the first 1--2\,Myr of the star-forming regions' evolution, which is shorter than the timescale for terrestrial planet formation.

This is helped by our simulated star-forming regions rapidly dissolving in the first few Myrs, placing the binary systems into the Galactic field. As such, our hard binaries typically only experience one hardening encounter. In less realistic models for the initial conditions of star forming regions (e.g. \citet{Plummer11} or \citet{King66} profiles), the star-forming reigons would be longer lived and so the binaries would experience further hardening or softening encounters.

Soft binary systems are highly likely to be disrupted in star-forming regions; however, habitable zones around the component stars in these systems would be distinct, and for our purposes would be treated as the habitable zones around single stars. In the dense star-forming regions we simulate hard binaries (5 -- 10\,au) are much more likely to be hardened, increasing the size of the habitable zone around both stars.\\

In summary, we find that dense stellar environments can enlarge the habitable zones around the components of close binary star systems, without detrimentally affecting the formation process for terrestrial planets that may reside in the habitable zone. 

\section*{Acknowledgements}

We are grateful to the referee, Maxwell Xu Cai, for helpful comments and suggestions which improved the original manuscript. We thank Stuart Littlefair, Chris Watson and Belinda Nicholson for helpful discussions. RJP acknowledges support from the Royal Society in the form of a Dorothy Hodgkin Fellowship.

\bibliographystyle{mnras}  
\bibliography{general_ref}

\label{lastpage}

\end{document}